# A Simplified Basis for Bell-Kochen-Specker Theorems

James D. Malley[a] and Arthur Fine[b]
[a]Center for Information Technology, National Institutes of Health, Bethesda, MD 20892, USA
[b]University of Washington, Seattle, WA 98195, USA

**Abstract**

We show that a reduced form of the structural requirements for deterministic hidden variables used in Bell-Kochen-Specker theorems is already sufficient for the no-go results. Those requirements are captured by the following principle: an observable takes a spectral value $x$ if and only if the spectral projector associated with $x$ takes the value 1. We show that the "only if" part of this condition suffices. The proof identifies an important structural feature behind the no-go results; namely, if *at least* one projector is assigned the value 1 in any resolution of the identity, then *at most* one is.

## 1. Introduction

The Bell-Kochen-Specker theorem targets noncontextual, deterministic hidden variables. John Bell [1] described the target this way: "The question at issue is whether the quantum mechanical states can be regarded as ensembles of states further specified by additional variables, such that given values of these variables together with the state vector determine precisely the results of individual measurements." Simon Kochen and Ernst Specker [2] spell out the requirement, implicit in Bell, that the measurement results determined by the additional variables must, on average, return the Born statistics of quantum theory for the state vectors in question. Yet Kochen and Specker do not regard even this requirement as sufficient, for they point to certain structural relations that quantum observables also satisfy, regardless of the state vectors, and they demand hidden variables satisfy these relations as well. The Bell-Kochen-Specker theorem, and its successors, show that a framework for hidden variables that respects these additional structural relations cannot reproduce the quantum statistics in any state space of dimension three or higher. For qubits, that is in dimension 2, Kochen-Specker, as well as Bell, show by example that all the requirements posed can be satisfied.

The structural requirements that Kochen and Specker point to, and that also drive Bell's work, can be formulated in several different ways. For Kochen and Specker the primary way involves values of functions of observables. For Bell it is the additivity of values for commuting observables. Similarly rules on values of products of commuting observables, or on joint distributions have also been considered. These requirements are roughly equivalent to the following simple restriction on values assigned by the hidden variables: an observable takes a spectral value x if and only if the spectral projector associated with x takes the value 1 [3, 4]. Given a standard framework for noncontextual deterministic hidden variables, we show that further weakening this requirement to the "only if" part – which we call *BKS/2* – is already sufficient for the Bell-Kochen-Specker no-go results.

**2. Deterministic Models**

We confine our discussion to discrete observables and treat a quantum system *S* with state space *H* in a state represented by a density operator *D*. The standard framework for noncontextual deterministic hidden variables associates with system *S* and state *D* a probability space $<\Lambda, \mu(.)>$ where the "hidden variables" are the elements in $\Lambda$, and $\mu(.)$ is a probability measure defined on a suitable $\sigma$ – algebra of subsets of $\Lambda$. The framework then associates each observable *A* on system *S* with a $\mu$ – measurable function *A*(.), often called a "response function," defined on $\Lambda$ and taking values in the spectrum of *A*. For $\lambda \in \Lambda$ we will write $A(\lambda) \doteq val_\lambda(A)$. Thus $val_\lambda(A)$ is always an eigenvalue of *A* and the response functions are random variables defined on $\Lambda$. The primary requirement on this framework is that the probability distribution of a random variable *A*(.) agree with the Born probabilities, $\Pr(A \in X | D)$, that a measurement of observable *A* in state *D* would yield an outcome in *X*. If $P_X$ is the projector onto the subspace spanned by eigenvectors of *A* with eigenvalues in *X*, $\Pr(A \in X | D) = tr(P_X D)$. Thus we require that

(1) $tr(P_X D) = \mu(A^{-1}(X)) = \mu(\{\lambda \in \Lambda | val_\lambda(A) \in X\})$.

Here the Born probabilities in state *D* for outcomes in *X* (on the left) match probabilities for values in *X* (the two terms on the right) determined by the hidden variables associated with *D*.

Note that for random variables *A*(.) and *B*(.) representing respectively, observables *A* and *B*, there is always a classical probability joint distribution, even if *A* and *B* do not commute. It is defined by $\Pr(A \in X \& B \in Y) = \mu(A^{-1}(X) \cap B^{-1}(Y))$ provided $A^{-1}(X)$ and $B^{-1}(Y)$ are $\mu$ – measurable. Using this a conditional probability $\Pr(A \in X | B \in Y)$ will be defined as well. Quantum joint probabilities $\Pr(A \in X \& B \in Y | D)$ in state *D* are already partially accounted for in this framework in the sense that, if *A* and *B* commute, then $\Pr(A \in X \& B \in Y | D) = \Pr(P_X P_Y = 1 | D)$. We see that for commuting *A*, *B* the product $P_X P_Y$ of two of their spectral projectors is also an observable. Therefore its distribution is already given in the hidden variables framework by (1) as $tr(P_X P_Y D)$, exactly as required by the Born rule.

Nevertheless, nothing up to now in this framework connects joint distributions of the random variables with the quantum joints and conditionals for the corresponding observables. A seemingly natural assumption would be to identify the quantum mechanical joint or conditional distributions, when defined, with the already defined joint or conditional distributions of the corresponding random variables. A distinguishing feature of these marginal and conditional distribution requirements is that they are experimentally testable. Further, precisely that identification, however arrived at, leads to the additional structural requirements already mentioned [1, 2]. If imposed for enough observables and states, using the joint distributions of the random variables this way suffices for virtually all the no-go theorems, including even the Bell Theorem [4] and Malley [5, 6], where the joint distribution structural requirements are shown to imply that all observables commute.

## 3. Main Results

In this section we show that a reduced form of the structural requirements used in Bell-Kochen-Specker theorems is already sufficient for the no-go results. We consider a state space **H** of dimension 3. The argument is easily generalized for spaces of higher dimension. Call an observable *maximal* if all its eigenspaces are just one dimensional. Then:

*Definition* (BKS/2). If $M$ is a maximal observable with spectral resolution $M = \sum_n x_n P_n$, then, for every $\lambda$, $val_\lambda(M) = x_n$ implies that $val_\lambda(P_n) = 1$.
(As explained in Sec.1 above, this corresponds to the "only if" part of the "if and only if" assumed by Bell-Kochen-Specker.)

Here and below *for almost all* $\lambda$ means for all but a set of $\lambda$-measure zero.

*Lemma* 1. BKS/2 implies that in every orthogonal triple $P_1, P_2, P_3$ of rank one projectors and for almost all $\lambda \in \Lambda$, $val_\lambda(P_m) = 1$, for *at least one m*.

*Proof.* Let $x_n \neq 0$ be distinct real numbers for $n = 1, 2, 3$ and for orthogonal projectors $P_n$, define a maximal observable $M$ by $M = \sum_n x_n P_n$. According to the Born rule the probability that a measurement of $M$ results in one of its eigenvalues is given by $tr(DP)$, where $P$ projects onto the subspace spanned by the eigenvectors of $M$. Because $M$ is maximal, that subspace is the whole three dimensional state space **H**. Thus $P$ is the identity on **H** and $tr(DP) = tr(D) = 1$. From (1), $\mu(M^{-1}(\{x_1, x_2, x_3\})) = 1$. So for almost all $\lambda$, it follows that $val_\lambda(M) = x_n$ for one or more $n \in \{1,2,3\}$ Hence from BKS/2, for almost all $\lambda$, $val_\lambda(P_m) = 1$, for at least one $m$. ∎

*Lemma* 2. If $P_1, P_2, P_3$ is an orthogonal triple of rank one projectors that resolves the identity and if $val_\lambda(P_m) = 1$, for at least one $m$, then the $val_\lambda(.)$ function is *orthogonally additive*; that is, exactly one projector $P_m$ has $val_\lambda(P_m) = 1$.
*Proof.* For $x, y, z \in \{0,1\}$ write
(2) $p(xyz) = \mu(\lambda \mid (val_\lambda(P_1), val_\lambda(P_2), val_\lambda(P_3)) = (x, y, z))$.
Since the triples $x, y, z$, partition the set of all possible values for the orthogonal triple $P_1, P_2, P_3$, the following total probability condition must hold:
(3) $1 = \sum_{x,y,z} p(xyz)$.

Since $tr[DP_1] = \Pr(P_1 = 1 | D)$, $tr[DP_2] = \Pr(P_2 = 1 | D)$, $tr[DP_3] = \Pr(P_3 = 1 | D)$, and the response functions return the Born probabilities,

(4) $tr[DP_1] = \sum_{y,z} p(1yz)$, $tr[DP_2] = \sum_{x,z} p(x1z)$, $tr[DP_3] = \sum_{x,y} p(xy1)$.

Since the orthogonal triple $P_1, P_2, P_3$ resolves the identity,

(5) $1 = tr[D] = tr[DP_1] + tr[DP_2] + tr[DP_3]$.

Thus

(6) $1 = \sum_{y,z} p(1yz) + \sum_{x,z} p(x1z) + \sum_{x,y} p(xy1)$

By assumption, the probability that $val_\lambda(P_j) = 0$ for each $j$ in an orthogonal triple is 0. Subtracting (3) from (6) yields

(7) $2p(111) + p(110) + p(101) + p(011) = 0$.

Since each $p(xyz) \geq 0$, all of the probabilities in (7) must be 0. Thus the only possible values for the response functions occurring with non-zero probability are

(8) $(val_\lambda(P_1), val_\lambda(P_2), val_\lambda(P_3)) \in \{(1,0,0),(0,1,0),(0,0,1)\}$. ∎

If we call any $val_\lambda(.)$ a *valuation*, then the lemmas imply the following result.

*Theorem.* If BKS/2 holds in any deterministic hidden variables model then there are valuations $val(.)$ satisfying:

(9a) $val(P) \in \{0,1\}$ for any projector $P$, and

(9b) if $P_1, P_2, P_3$ is an orthogonal triple of rank one projectors that resolves the identity, then exactly one projector $P_m$ has $val(P_m) = 1$.

These conditions define a frame function of weight 1, in violation of Gleason's theorem [7], and they are the starting conditions that Bell [1] uses to prove a violation of the quantum statistics for a finite number of projectors. They are also the basis of the similar no-go proved by Kochen and Specker [2] as well as the theorem of Malley [8] demonstrating that the ambient space $H$, which was assumed to be of dimension 3, cannot be of dimension 3. (Effectively, from within the model, the ambient space $H$ looks like a single projector surrounded by other projectors orthogonal to it.)

**4. Joint Distributions and BKS/2**

To appreciate the work done by BKS/2 it may be useful to see how quickly the theorem here falls out of assumptions concerning joint distributions; namely, that quantum joint distributions of observables, when defined, coincide with the *always defined* joint distributions of the random variables that represent those observables. Consider then an orthogonal triple $P_1, P_2, P_3$ of rank one projectors that resolves the identity. Since $P_i P_j = 0$ for i≠j, the Born probability is 0 that any two projectors both have value 1. Then, given the stated assumption, for i≠j the distributions of the corresponding random variables yield that $\mu(\{\lambda | val_\lambda(P_i) = 1 \ \& \ val_\lambda(P_j) = 1\}) = 0$. Hence for almost all $\lambda$ no more than one of the projectors $P_m$ has $val_\lambda(P_m) = 1$. But since the Born probability that all three have value 0 is also 0, the same must be true of the corresponding random variables; that is, $p(000) = 0$. Hence for almost all $\lambda$ we have $val_\lambda(P_m) = 1$ for exactly one $m$. Thus equations (9a) and (9b) are satisfied, and the *Theorem* holds. Note that we only need to use the joint distributions of the random variables to establish that $p(000) = 0$, since the result then follows directly from *Lemma* 2.

**5. Discussion**

Our results reveal a new, minimal structure behind the now classical no-go theorems of the Bell-Kochen-Specker type. Lemma 2 shows that in a standard framework for deterministic hidden variables if *at least* one projector is assigned the value 1 in any resolution of the identity, then *at most* one is. Hence if at least one projector is assigned the value 1, the assignment induced by the hidden variables is orthogonally additive and the no-go theorems follow from the geometry of the state space alone without further assumptions. What underlies this structure is the assumption that values assigned by the hidden variables to an observable always lie in its spectrum. By contrast, for observables other than position, this is not true in the de Broglie-Bohm introduction of hidden variables [9], nor in any approach that includes null outcomes among the responses due to hidden variables [10, 11]. Indeed both these ways of using hidden variables avoid the classical no-go theorems.

Recent no-go theorems have been set in what is called an "ontological" framework for hidden variables [12]. This includes an attempt by Pusey, Barrett and Rudolph (PBR) to eliminate an "epistemic" interpretation of the wave function in favor of a "realist" one [13], and a very broad no-go theorem that challenges the composition principle at the heart of the PBR result [14]. In the deterministic case the ontological framework reduces to the standard framework studied here, and the broad theorem in question uses an assumption ("Assumption A") that, restricted to maximal observables, is what we have, above (Sec. 3), called BKS/2. Our study shows that BKS/2 is sufficient for orthogonal additivity. Thus, provided the state space is of dimension 3 or higher, BKS/2 alone leads to no-go results of the Bell-Kochen-Specker type. The no-go theorem in [14], however, despite its broad scope, only requires BKS/2 for qubits, where it is known to be consistent and harmless. Thus the connection we show of BKS/2 with orthogonal additivity does not entail the no-go results of that theorem, nor undermine the challenge it poses to the PBR composition principle.

Finally, given the minimal requirements for satisfying Lemma 2, it may be worth examining other versions of the standard assumptions used in the no-go theorems to see whether, as in BKS/2, they too can be weakened. Weaker assumptions make for stronger theorems. They may also impact schemes for simulating quantum effects classically, a topic of interest in several applied areas [15, 16, 17].

**Acknowledgement**  JDM acknowledges support from the NIH Intramural Research Program. JDM and AF want to thank Maximilian Schlosshauer for his careful reading of the manuscript and for identifying important links to the applied literature.